# High spatial coherence and short pulse duration revealed by the Hanbury Brown and Twiss interferometry at the European XFEL

RUSLAN KHUBBUTDINOV,[1,2] NATALIA GERASIMOVA,[3] GIUSEPPE MERCURIO,[3] DAMELI ASSALAUOVA,[1] JEROME CARNIS,[1] LUCA GELISIO,[4] LOÏC LE GUYADER,[3] ALEXANDER IGNATENKO,[1] YOUNG YONG KIM,[1] BENJAMIN VAN KUIKEN,[3] RUSLAN P. KURTA,[3] DMITRY LAPKIN,[1] MARTIN TEICHMANN,[3] ALEXANDER YAROSLAVTSEV,[3] OLEG GOROBTSOV,[5] ALEKSEY P. MENUSHENKOV,[2] MATTHIAS SCHOLZ,[3] ANDREAS SCHERZ,[3] AND IVAN A. VARTANYANTS [1,2,*]

[1] *Deutsches Elektronen-Synchrotron DESY, Notkestraße 85, D-22607 Hamburg, Germany*
[2] *National Research Nuclear University MEPhI (Moscow Engineering Physics Institute), Kashirskoe shosse 31, 115409 Moscow, Russia*
[3] *European XFEL, Holzkoppel 4, 22869 Schenefeld, Germany*
[4] *Center for Free-Electron Laser Science, DESY, Luruper Chaussee 149, D-22761 Hamburg, Germany*
[5] *Department of Materials Science and Engineering, Cornell University, Ithaca, New York 14850, USA*
*\* Ivan.Vartaniants@desy.de*

**Abstract:** Second-order intensity interferometry was employed to study the spatial and temporal properties of the European X-ray Free-Electron Laser (EuXFEL). Measurements were performed at the soft X-ray SASE3 undulator beamline at a photon energy of 1.2 keV in the Self-Amplified Spontaneous Emission (SASE) mode. Two high-power regimes of the SASE3 undulator settings, *i.e.* linear and quadratic tapering at saturation, were studied in detail and compared with the linear gain regime. The statistical analysis showed an exceptionally high degree of spatial coherence up to 90% for the linear undulator tapering. Analysis of the measured data in spectral and spatial domains provided an average pulse duration of about 10 fs in our measurements. The obtained results will be valuable for the experiments requiring and exploiting short pulse duration and utilizing high coherence properties of the EuXFEL.

## 1. Introduction

With the development of pulsed lasers in the visible spectrum and the invention of X-ray Free-Electron Laser (XFEL) sources in the X-ray energy range [1], it has become clear that for a vast majority of experiments essential information about the sample can only be deduced from multiple measurements. This may be accomplished by many realizations of the radiation field, *i.e.* multiple pulses, from these sources. For example, in single particle imaging experiments performed at XFEL sources [2], hundreds of thousands of coherent X-ray pulses interact with differently oriented replicas of the target particle in order to determine its three-dimensional structure with high resolution [3-5]. It is also clear that these pulsed sources, in principle, cannot be treated as stationary ones and, hence, such properties as spatial and temporal coherence have to be revised [6].

Coherence, at its basics, is the manifestation of correlations of the optical wave fields [7,8]. The first order coherence or correlations of the field amplitudes may be experimentally probed by Young's double pinhole or Michelson split-and-delay type of experiments [7,8]. Another important approach to tackle coherence is to perform the second-order intensity correlation measurements, originally proposed by Hanbury Brown and Twiss (HBT) in their pioneering experiments [9,10]. Importantly, these experiments led to the creation of the field of quantum optics [11,12] and became pivotal for the realization of quantum imaging and quantum technology [13-16].

The basic idea of HBT interferometry is to explore correlation of intensities at different spatial or temporal positions, i.e. to perform measurements of the second-order correlation functions. For example, if measurements are carried out in spatial domain, this leads to the following normalized second-order correlation function

$$g^{(2)}(x_1, x_2) = \frac{<I(x_1)I(x_2)>}{<I(x_1)><I(x_2)>}, \qquad (1)$$

where $I(x_1), I(x_2)$ are the intensities of the wave field and averaging, denoted by brackets <...>, is performed over a large ensemble of different realizations of the wave field. If radiation is cross-spectrally pure and obeys Gaussian statistics [8], which is typical for chaotic fields, then $g^{(2)}$-function can be expressed as [17-19]

$$g^{(2)}(x_1, x_2) = 1 + \zeta(D_\omega) \cdot \left|g^{(1)}(x_1, x_2)\right|^2, \qquad (2)$$

where $g^{(1)}(x_1, x_2) = <E^*(x_1)E(x_2)>/\sqrt{I(x_1)I(x_2)}$ is the first-order correlation function and $\zeta(D_\omega)$ is the contrast function, which depends on the radiation bandwidth $D_\omega$. In the limit when an average pulse duration $T$ is much larger than the coherence time $\tau_c$ ($T \gg \tau_c$), the contrast function is given by $\zeta(D_\omega) = \tau_c/T$ or inverse number of modes $M_t$ in temporal domain. In the opposite limit $T \ll \tau_c$ the contrast function reaches a constant value [17-19].

It is especially important to understand the coherence properties of recently constructed hard XFEL facilities [20-23] as many experiments rely on the high degree of coherence of these sources. The first-order coherence properties of these sources were determined in double pinhole Young's experiments [24,25] or by the speckle contrast analysis [26-28]. By performing HBT experiments at XFEL sources, rich information on their statistical properties, such as the degree of spatial coherence and average pulse duration can be determined [18,29-31] (see for review [32]). These experiments may also shed light on the fundamental statistical properties of XFEL sources and clearly indicate whether an FEL behaves as a true single-mode laser source or rather as a chaotic source of radiation [33]. Here, we present a statistical analysis of experimental results for the high-power European XFEL (EuXFEL) radiation by means of HBT interferometry. We performed experiments in both the spectral and spatial domains and characterized three XFEL operation modes: linear (LT) and quadratic (QT) undulator tapering at saturation, as well as the operation in the linear gain regime (LR). Finally, we determined the degree of coherence and characteristic pulse durations for all three regimes.

## 2. Experiment

The experiment was performed at the Spectroscopy and Coherent Scattering (SCS) instrument of the European XFEL [34]. The EuXFEL linac was operated at 14 GeV electron energy with an electron bunch charge of 250 pC in a single bunch mode at 10 Hz repetition rate. The SCS instrument is located at the SASE3 undulator beamline which produces intense X-ray pulses in the soft X-ray photon energy range (250 eV - 3000 eV). A schematic representation of the beamline is shown in Fig. 1. The experiment was performed at 1.2 keV photon energy with three undulator configurations, i.e. LT and QT at saturation which are dedicated settings to provide high power radiation, and a configuration for operating the XFEL in the LR with the average pulse energies of 1.2 mJ, 6.5 mJ, and 0.117 mJ, respectively.

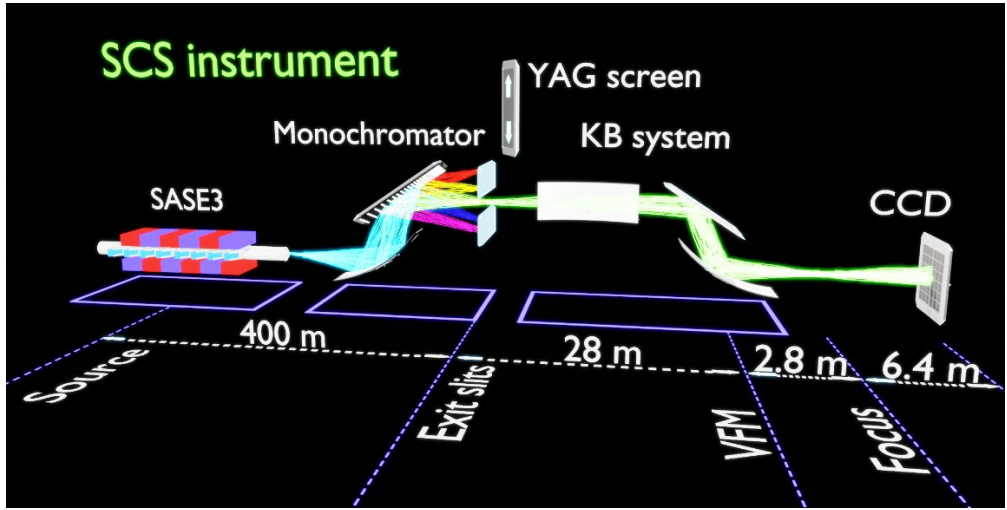

**Fig. 1.** Schematic layout of the experiment. The monochromator focuses the SASE3 undulator source located about 300 m upstream onto the exit slits (ES) located 100 m downstream. Vertical focusing mirror (VFM), part of a variable bending mirrors Kirkpatrick-Baez (KB) system, located 28 m downstream from the ES, refocuses the X-ray beam at 2.8 m downstream from this mirror. The measurements were taken at a distance of about 6.4 m from the focal position. Horizontal KB focusing mirror located 1.35 m upstream of the VFM, refocuses the intermediate horizontal focus located at 374 m after the undulator source.

The SCS instrument is equipped with a variable line spacing (VLS) diffraction grating monochromator [35,36]. The monochromator was operating in the second diffraction order, corresponding to a photon energy dispersion of 2.69 eV/mm (at 1.2 keV) in the exit slit (ES) plane along the vertical direction. The experimental resolution at this energy was estimated to be better than 0.3 eV at the full-width at half maximum (FWHM) (theoretical resolution at this energy is 0.2 eV [36]). In the case of spectral measurements, the spectral distribution of XFEL pulses were acquired in the ES plane of the monochromator. This was achieved by introducing a YAG:Ce crystal just behind the fully opened ES and detecting the optical luminescence from this crystal by a charge coupled device (CCD) gated by a microchannel plate (MCP).

Spatial second-order correlation measurements were performed by a back-illuminated CCD detector (Andor iKon-M 934, 1024×1024 pixels, pixel size of 13×13 μm$^2$) with 8 pixels binned in the vertical direction and no binning in the horizontal direction. The detector was located at the end of the beamline at a distance of about 6.4 m from the X-ray beam focus. The bending of the horizontal Kirkpatrick-Baez (KB) mirror was adjusted for QT and LT/LR measurements, thus changing the horizontal size of the beam on the detector (Fig. 1). In the case of spatial correlation measurements, the bandwidth of X-ray radiation at the CCD detector was controlled by the size of the monochromator ES opening.

## 3. Results

### 3.1 Spectral analysis

Single-pulse spectra were recorded for all three regimes of the EuXFEL operation prior to spatial measurements. Each run consisted of about $3 \cdot 10^3$ pulses for each operating condition of the EuXFEL. Below we present the results of spectral and spatial measurements for the LT regime. As shown in Fig. 2(a), one can observe a multimodal structure in the single pulse spectrum intensity distribution for all operating regimes of the EuXFEL. The number of spectral modes varies depending on the operation conditions. As it is well seen in Fig. 2(a), the average spectrum does not resemble a single Gaussian function but is rather a sum of two distributions, which applies also to the other studied operating conditions. We estimated the FWHM of the average normalized spectrum directly from the experimental data and determined that it was in the range of 0.7% - 1% from the resonant energy, depending on the experimental conditions (see Fig. 2(a) and Table 1).

This spectrum is about three times wider than the theoretically predicted one for the SASE3 undulator [37], which may be affected by the energy chirp of the beam. The knowledge of the average spectrum allows one to estimate the coherence time of X-ray radiation as

$$\tau_c = \int_{-\infty}^{\infty} |\gamma(\tau)|^2 d\tau, \qquad (3)$$

where $\gamma(\tau)$ is the complex degree of coherence [7,8]. The complex degree of coherence may be determined through the average spectral density $S(\omega)$ as

$$\gamma(\tau) = \frac{\int_0^{\infty} S(\omega) e^{-i\omega\tau} d\omega}{\int_0^{\infty} S(\omega) d\omega}. \qquad (4)$$

We fitted the average spectrum by the sum of two Gaussian functions and determined in this way the coherence time in each operating condition of the SASE3 undulator (see Table 1). For all three operation conditions the coherence time was in the range from 200 as to 300 as.

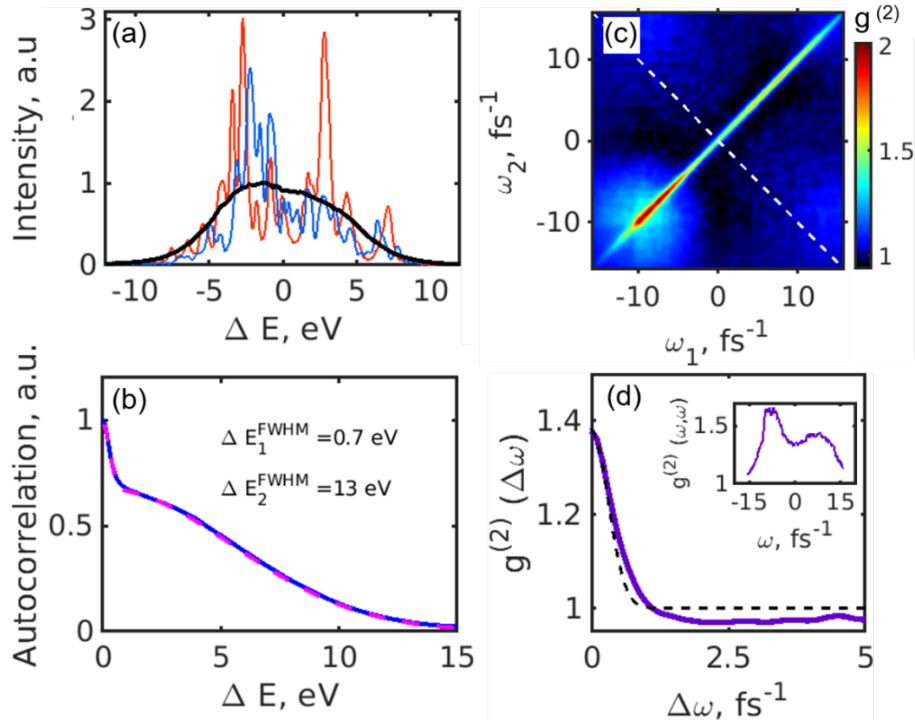

**Fig. 2.** (a) Typical single shot spectra (red and blue lines) for the LT operation regime and an average spectrum for about $3 \cdot 10^3$ pulses (black line). (b) Autocorrelation function of individual spectral lines averaged over the same number of pulses (blue solid line) and its fit with a sum of two Gaussian functions (magenta dashed line). The values of the FWHMs of these two Gaussian fits is shown in panel (b). (c) Second-order intensity correlation function of spectra $g^{(2)}(\omega_1, \omega_2)$. (d) Cut along the diagonal line shown by the white dashed line in (c) and its fit (black dashed line). In the inset the profile along the diagonal of the $g^{(2)}(\omega_1, \omega_2)$-function, with $\omega_1 = \omega_2$, in (c) is shown.

To determine the bandwidth of single spike spectral lines we performed the autocorrelation analysis of the spectra. The result of this analysis for the LT is shown in Fig. 2(b). For all three undulator settings, we observed similar features in the autocorrelation spectrum, *i.e.* a sharp peak (corresponding to the FEL spike width) standing on the pedestal of a broad peak (corresponding to the averaged FEL bandwidth). The values of the FWHM of both peaks, corrected for the factor of $\sqrt{2}$, are provided in Table 1.

The average pulse duration in spectral domain was determined by performing correlation analysis [19]. The second-order correlation function in the frequency domain in the case of the LT operation is presented in Fig. 2(c).

In Fig. 2(d) we analyzed its behavior along the white dashed diagonal line shown in Fig. 2(c). Note, that the contrast value is below unity for all three undulator settings due to the finite values of the degree of spatial coherence and monochromator resolution function (see Ref. [19]). Fitting this profile with a function [19]

$$g_{in}(\omega_1, \omega_2) = \frac{exp\left[-\frac{\sigma_T^2}{1 + 4\sigma_r^2\sigma_T^2}(\omega_2 - \omega_1)^2\right]}{(1 + 4\sigma_r^2\sigma_T^2)^{1/2}}, \quad (5)$$

which considers monochromator resolution $\sigma_r$, gave us the root mean square (rms) values $\sigma_T$ of an average pulse duration. From these values we deduced the FWHM values T=2.355·$\sigma_T$ to be in the range from 9 fs to 20 fs for all three undulator settings (see Table 1). The largest value of 20 fs was obtained for the QT regime of operation. We should note here that for high power radiation the YAG:Ce crystal may produce non-linear response and, hence, reduce the measured contrast values. This can affect our estimates of pulse duration for the QT regime of operation. These effects should not have any impact in the HBT measurements performed in spatial domain.

**Table 1. Results of the analysis in spectral domain**

| Operation regime | LT | QT | LR |
|---|---|---|---|
| Photon energy, eV | 1205 | 1205 | 1202 |
| FEL bandwidth (FWHM), eV | 10.0±0.1 | 12.6±0.1 | 8.8±0.1 |
| Width of spectrum from autocorrelation (FWHM), eV | 9.2±0.1 | 12.0±0.1 | 8.5±0.1 |
| Width of spike spectral lines from autocorrelation (FWHM), eV | 0.49±0.02 | 0.49±0.02 | 0.57±0.02 |
| Coherence time (rms), as | 309±7 | 235±5 | 316±7 |
| Pulse duration T from HBT spectral measurements (FWHM), fs | 13.0±0.2 | 20.0±0.4 | 8.7±0.1 |

## 3.2 Spatial analysis

Typical intensity distributions of individual pulses, measured with small and wide ES width, are shown in Fig. 3(a,b) for the LT mode. A visual inspection of the individual pulses reveals that between one and five spectral spikes were present in the XFEL beam at 2.5 mm wide ES. For further intensity correlation analysis, these intensity distributions were projected along the vertical (dispersion) direction and correlation analysis was performed in the horizontal direction according to Eq. (1) for about $10^4$ XFEL pulses.

**Table 2. Results of the analysis in spatial domain**

| Operation regime | LT | QT | LR |
|---|---|---|---|
| Average beam size (FWHM), mm | 2.3±0.03 | 1.7±0.02 | 2.2±0.03 |
| Coherence length (rms), mm | 2.4±0.1 | 1.3±0.4 | 2.6±0.2 |
| Degree of spatial coherence, % | 91.6±3 | 71±5 | 89±3 |
| Pulse duration T from HBT spatial measurements (FWHM), fs | 8.5±1.1 | 12.8±1.5 | 7.3±1.2 |

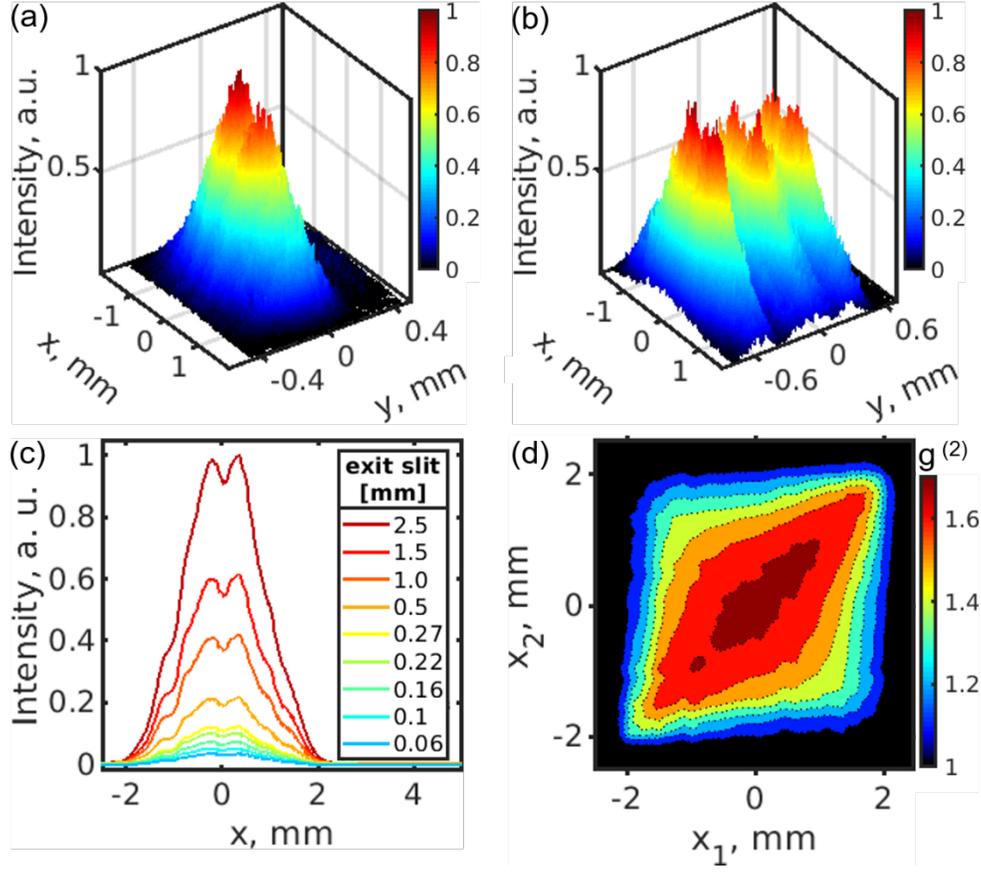

**Fig. 3.** (a,b) Typical intensity distributions of single pulses in the LT mode measured by the CCD detector for the ES width of 0.02 mm (a) and 2.5 mm (b). (c) Pulse intensities averaged along the vertical (dispersion) direction as a function of the ES width. The top curve corresponds to the ES width of 2.5 mm and the lowest one to the ES width of 0.06 mm. (d) Second-order intensity correlation function $g^{(2)}(x_1, x_2)$ for the ES width of 0.02 mm.

The average pulse intensity distribution for different monochromator ES width and typical intensity correlation $g^{(2)}$-function for the smallest ES width of 0.02 mm in the LT mode are shown in Fig. 3(c,d). We clearly observe the growth of the average pulse intensity with the increase of the monochromator ES width. The non-Gaussian shape of these intensity profiles we attribute to the slope error of the beam transport mirrors. Directly from these profiles we estimated the FWHM of the beam size to be in the range from 1.7 mm to 2.3 mm, depending on the KB mirror settings and mode of the undulator operation (see Table 2). As it can be seen from Fig. 3(d) the shape of the $g^{(2)}$-function resembles a flat-top function. Such form of the $g^{(2)}$-function is typical for highly coherent radiation when the coherence length of radiation is much larger than the size of the beam [30,38].

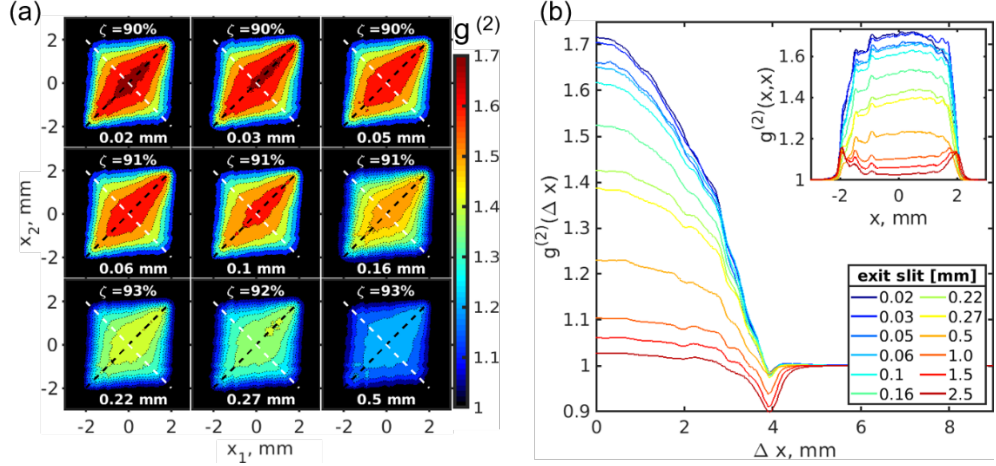

**Fig. 4.** (a) Second-order intensity correlation functions $g^{(2)}(x_1, x_2)$ determined in measurements in the LT mode. Each panel corresponds to a certain width of the monochromator ES which is indicated in the panel. (b) Profiles of the $g^{(2)}(\Delta x)$-function taken along the white dashed antidiagonal lines shown in panel (a) as a function of the ES width. In the inset the corresponding $g^{(2)}(x,x)$-functions taken along the black dashed diagonal lines in panel (a) are shown.

In Fig. 4(a) the results of intensity correlation analysis for the LT mode as a function of the monochromator ES width are presented. The intensity correlation functions determined along the white dashed lines in Fig. 4(a) are presented in Fig. 4(b) for different ES width. One can see that the larger the ES width the lower is the contrast $\zeta(D_\omega)$ in Eq. (2) (that is the maximum value of the $g^{(2)}(\Delta x)$ function), which obeys the typical behavior predicted by this equation for a Gaussian chaotic source. We also notice that for the separation of two points of about $\Delta x = 4$ mm, the correlation function $g^{(2)}(\Delta x)$ takes values below unity. We attribute these features to low values of intensity at these separations and to positional jitter of the beam [29].

Next, we determined the degree of spatial coherence $\zeta$ and the coherence length $L_{coh}$ as a function of coherence time for all three undulator settings (see Fig. 5). The coherence length $L_{coh}$ was obtained as the variance value of the $g^{(2)}(\Delta x) - 1$ function for the different ES width. The coherence time was determined according to Eqs. (3,4), in which the spectral density $S(\omega)$ was substituted by the function $\tilde{T}_{sl}(\omega) = T_{sl}^2(\omega) \otimes R(\omega)$. This function accounts for the finite monochromator resolution function $R(\omega)$ and considers the slit function $T_{sl}(\omega)$ in the form of the rectangular function [19]. The first observation here is that the degree of spatial coherence and coherence length essentially do not depend on the coherence time, being practically constant in the range of coherence times from 1 fs to 12.8 fs. Second, one can see that the value of coherence length for the LT and LR settings was about the FWHM width of the corresponding beams. For the QT we observed slightly smaller values of the coherence length, which were also lower than the corresponding beam sizes (see Table 2). From these observations we concluded that the degree of spatial coherence reaches high values of about 85-95% for the LT and LR modes. In the case of QT mode we observed a slightly lower degree of spatial coherence of about 70% [39].

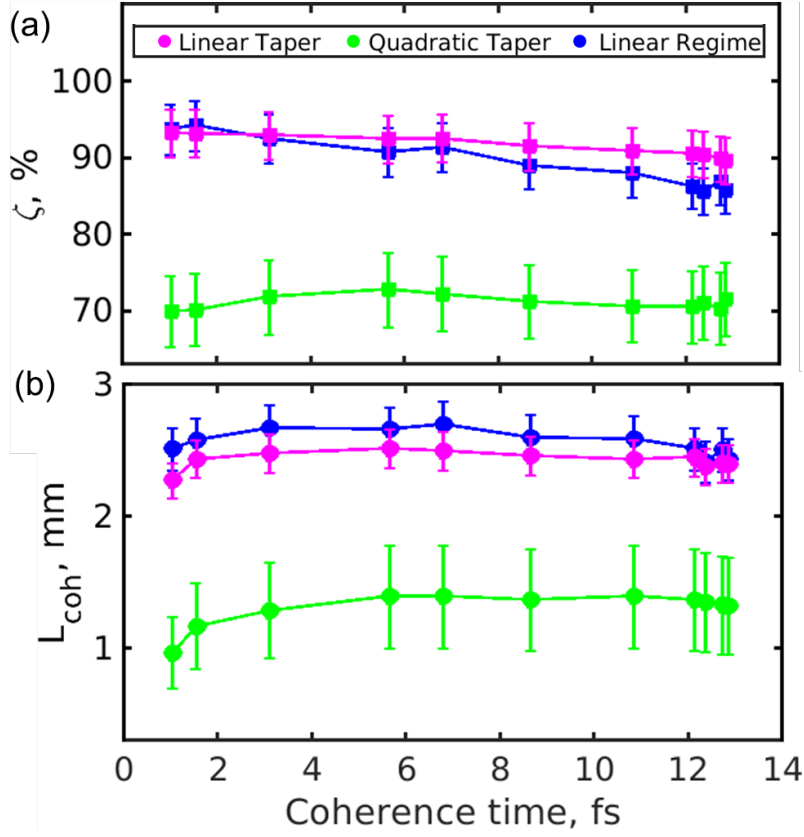

**Fig. 5.** Degree of coherence (a) and coherence length (b) as a function of coherence time for different undulator operation modes.

We also determined the values of contrast $\zeta(D_\omega) = g^{(2)}(x,x) - 1$ taken at $x = 0$ as a function of coherence time for all three settings of the undulators. As it is clearly visible from Fig. 6 the values of contrast strongly depend on coherence time and show the typical behavior of a chaotic source [18,29]. One may observe in Fig. 6 that the contrast values do not reach the maximum value of unity at saturation, which is due to the finite resolution of the monochromator (see Ref. [19]). The obtained contrast values were fitted with the following equation

$$\zeta_{in}(D_\omega) = \frac{\int_{-\infty}^{\infty} F(\omega) \left|g_{in}^{(1)}(\omega)\right|^2 d\omega}{\left[\int_{-\infty}^{\infty} \tilde{\tilde{T}}(\omega) d\omega\right]^2}, \qquad (6)$$

where $F(\omega)$ is the autocorrelation function

$$F(\omega) = \int_{-\infty}^{\infty} \tilde{\tilde{T}}(\omega') \tilde{\tilde{T}}(\omega' + \omega) \, d\omega' \qquad (7)$$

and the function $\tilde{\tilde{T}}(\omega)$ is defined as

$$\tilde{\tilde{T}}(\omega) = S_{in}(\omega) \tilde{T}_{sl}(\omega). \qquad (8)$$

In Eq. (6) the function $g^{(1)}_{in}(\omega)$ is defined as

$$g^{(1)}_{in}(\omega) = exp\left[-\frac{\sigma_T^2 \omega^2}{2}\right]. \qquad (9)$$

This fitting provided us with the average pulse duration values from about 7 fs to 13 fs for the different SASE3 undulator settings of the EuXFEL (see Table 2). Here, as before, we assumed the resolution value of the monochromator $R(\omega)$ to be 0.3 eV.

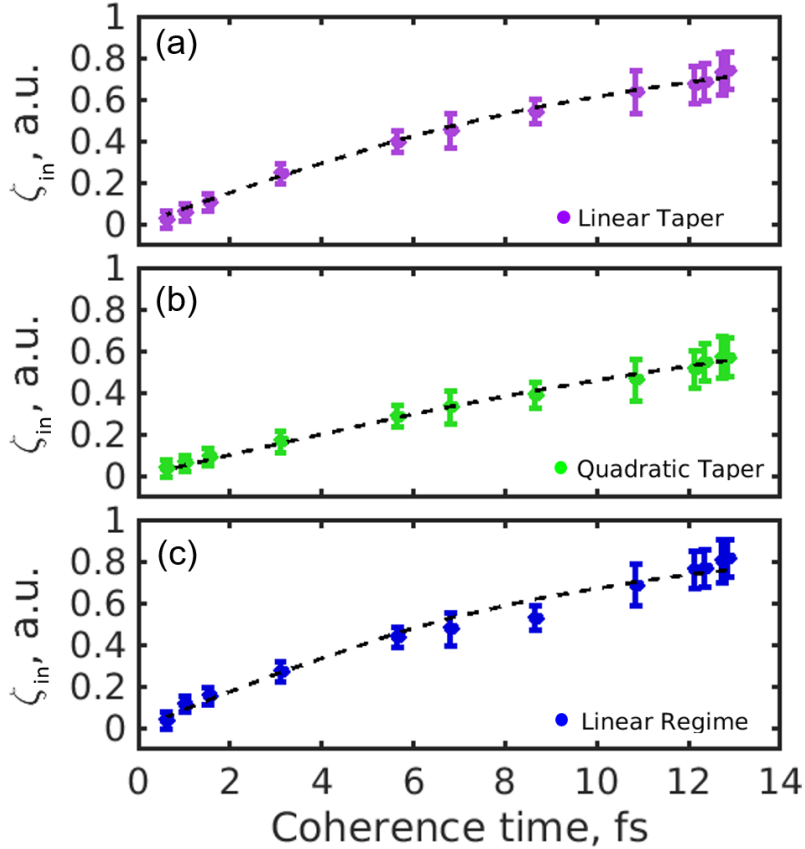

**Fig. 6.** Contrast values as a function of coherence time for different modes of undulator settings. (a) LT, (b) QT, and (c) LR of the undulator operation. Circles represent experimental points. Errors are calculated as the standard deviation of the contrast. The black dashed lines are the fit over all experimental data points.

## 4. Discussion and summary

The second-order correlation experiment performed at the high-power SASE3 undulator of the EuXFEL facility at 1.2 keV photon energy demonstrated the high degree of spatial coherence of radiation using the LT (90%) and QT (70%) undulator settings at saturation, as well as operating in the LR (90%). These values are consistent with the results of experiments performed at different XFEL sources but show a higher degree of spatial coherence at the EuXFEL in comparison to the other XFEL sources [18,24-30]. This may be attributed to the higher electron energy operation of 14 GeV of EuXFEL as suggested in Ref. [39].

By performing HBT interferometry it was possible to determine not only the degree of spatial coherence, but also the average pulse duration of radiation before the monochromator. The determined values of the pulse duration both in spectral and spatial domains were on the order of 10 fs (except of QT mode of operation) (see Tables 1 and 2). These short pulse durations should be considered with the certain caution. The nominal pulse duration of the European XFEL was about 20 fs as determined from bunch length for the electron bunch measurements [23]. The spectral bandwidth of the SASE3 undulator in our experiment was ~1% and thus about 2-3 times larger than the baseline parameter, i.e. 0.35% at 1200 eV [37]. The observed broadening of the spectral profiles including individual spikes may be caused by the finite monochromator resolution as well as frequency chirp of the X-ray pulses. As soon as the broadening due to monochromator resolution is about 3%, we attribute most of the observed broadening to the frequency chirp of the X-ray pulses, which is a result of the electron bunch chirp in the accelerator modules. As it follows from our simulations, the spectral width, as well as the spectral spikes width, may change significantly due to the frequency chirp effects. This may cause an apparent lower pulse duration obtained from both spectral and spatial measurements from the nominal one. Thus, pulse durations upstream of the monochromator can be about twice longer than deduced from our analysis, lying in the range of 10 fs to 20 fs. As such, our measurements provide the lower boundary for the pulse durations of the European XFEL at different modes of operation.

From this discussion it is clear that additional measurements with a controlled frequency chirp of radiation will be an important step in understanding the properties of the EuXFEL radiation. It will be also important to carry out measurements using complementary methods, for example, gas ionization at the same photon energy of 1.2 keV at the Small Quantum Systems (SQS) instrument [34] that is sharing the same undulator SASE3 at the EuXFEL facility.

In summary, the statistical analysis of X-ray radiation by means of HBT interferometry is a powerful tool to understand the basic properties of the beams generated by the soft X-ray undulators at the EuXFEL. Such vital parameters as the degree of coherence and pulse duration can be determined in these experiments. The results obtained in this work will be of extreme importance for the experiments requiring and utilizing the high coherence properties and short pulse durations of the EuXFEL.

**Acknowledgements**

We acknowledge European XFEL in Schenefeld, Germany, for provision of X-ray free-electron laser beamtime at Scientific Instrument SCS (Spectroscopy and Coherent Scattering) and would like to thank the staff for their assistance. The authors acknowledge support of the project by E. Weckert and careful reading of the manuscript by M. Beye.

**Funding**

The authors acknowledge financial support from the Helmholtz Associations Initiative Networking Fund (grant No. HRSF-0002) and the Russian Science Foundation (grant No. 18-41-06001). A. Ya. and A.P.M. acknowledge the Russian Foundation for the Basic Research (grant No. 18-02-40001 mega) for financial support.